 \documentclass[final,5p,times,twocolumn]{elsarticle}

\usepackage{amssymb}

\usepackage{ascmac}
\usepackage{amsmath}
\newcommand{\del}{\partial}

\journal{Physics Letters B}

\makeatletter
\def\ps@pprintTitle{%
  \let\@oddhead\@empty
  \let\@evenhead\@empty
  \let\@oddfoot\@empty
  \let\@evenfoot\@oddfoot
}
\makeatother

\begin{document}

\begin{frontmatter}



\title{On Geometries and Monodromies for Branes of Codimension Two}


\author[first]{Tetsuji Kimura}
\author[second]{Shin Sasaki}
\author[second]{Kenta Shiozawa}

\affiliation[first]{organization={Center for Physics and Mathematics,
 Institute for Liberal Arts and Sciences, \\
Osaka Electro-Communication University},
            city={Neyagawa},
            postcode={572-8530}, 
            state={Osaka},
            country={Japan}}

\affiliation[second]{organization={Department of Physics, Kitasato University},
            postcode={252-0373}, 
            state={Sagamihara},
            country={Japan}}

\begin{abstract}
We study geometries for the NS5-, the KK5- and the $5^2_2$-branes of codimension two 
in type II and heterotic string theories.
The geometries are classified by monodromies that each brane has.
They are the $B$-, the general coordinate and the $\beta$-transformations of the
 spacetime metric, the $B$-field and the dilaton (and the gauge fields).
We show that the monodromy nature appears
 also in the geometric 
quantities
 such as the curvature and the complex structures of spacetime.
They are linearly realized in the doubled (generalized) structures in the doubled space.
\end{abstract}




\end{frontmatter}


\section{Introduction} \label{sec:introduction}

The background geometries for branes of codimension two are
characterized by monodromies \cite{Vafa:1996xn, deBoer:2012ma}.
This property is particularly associated with 
string dualities.
T-folds, which are spacetime structures unique to string theory, 
emerge from T-duality transformations of conventional geometry 
\cite{Hull:2004in, Hull:2006va}.
The T-folds locally look like Riemann(-Cartan) spaces, but globally they are not.
They are defined as spacetimes where the fields on them undergo the monodromy of a $\beta$-transformation.
On the other hand, there are geometric quantities that characterize
spacetimes, such as connection, curvature, torsion, and so on. 
In the T-fold spacetimes, these geometric quantities are no longer 
singlevalued
 functions of spacetime and are not only patched by the general coordinate transformation, 
but also by the T-duality transformation.
Thus, they lose their geometric meaning in the usual sense.

In this letter, we examine appropriate geometric quantities that describe T-folds.
As a prototypical example, we consider the NS5-, the KK5-, and the
$5^2_2$-branes of codimension two in type II and heterotic theories, and study their geometric structures.
These geometries are classified by monodromies.
The NS5- and the KK5-branes have monodromies associated with the
$B$-transformation and the general coordinate transformation, hence they are geometric.
On the other hand, the $5^2_2$-brane has a monodromy of the
$\beta$-transformation which is non-geometric.
This property becomes apparent by viewing the corresponding geometries in
the doubled (generalized) language in $2D$ dimensions.
We show that the monodromy nature carries over to the generalized
connection, the generalized curvature and the generalized complex
structures in the doubled space.
In our example, the T-fold is formulated as a 
$2D$-dimensional doubled complex manifold $\mathcal{M}$ with the intrinsic
monodromy of the $\beta$-transformation.
This doubled manifold $\mathcal{M}$ has an extended Born structure
\cite{Kimura:2022jyp}, and all the geometric quantities
are defined up to the monodromy of the $\beta$-transformation.

The organization of this letter is as follows.
In the next section, we briefly introduce the $D$-dimensional spacetime geometries for
five-branes of codimension two.
In section \ref{sec:doubled_formalism}, we introduce the doubled
formalism and discuss the monodromy of the complex structures.
In section \ref{sec:codim_two_connection}, we study the connections and
curvatures of the geometries in the doubled formalism and show that the
monodromy of the $\beta$-transformation appears also in these quantities.
Section \ref{sec:conclusion} is devoted to the conclusion and discussions.

\section{Geometries for five-branes of codimension two}
We first look at the geometry of the NS5-brane of codimension two.
This is a solution to type II supergravities and given by
\begin{align}
&
ds^2 = H (r) (d\vec{x})^2,
\qquad
B = A \, dx^8 \wedge dx^9,
\qquad
e^{2\phi} = H (r), 
\notag \\
&
H (r) = h_0 + \sigma \log \frac{\mu}{r},
\qquad 
A = - \sigma \theta,
\notag \\
& 
x^6 = r \cos \theta, \quad x^7 = r \sin \theta,
\label{eq:NS5_solution}
\end{align}
where
the spacetime metric is in the string frame and 
$h_0, \sigma, \mu$ are constants.
Here we consider only the four-dimensional transverse directions 
$\vec{x} = (x^6,x^7,x^8,x^9)$ while the worldvolume directions $(x^0, \ldots,
x^5)$ have been omitted.
The spacetime metric $g_{\mu \nu}$ and the dilaton $\phi$ do not depend on $\theta$, while 
$B$ linearly depends on $\theta$.
The fact that the $B$-field depends only on $\theta$ in the base space
$(x^6,x^7)$ is a feature of codimension two solution. 
Then the monodromy by $\theta = 0 \to 2 \pi$ is represented as a shift
of the $B$-field.
When $(g,B, \phi)$ is subject to the factorized T-duality
transformation $T_k$ along the $x^k$-direction, 
the $\theta$-dependence in the $B$-field appears 
as a part of
 components in the
spacetime metric.
For example, the $5^2_2$-brane solution, which is given by the $T_8
T_9$-transformations of the NS5-brane, is given by
\begin{align}
& 
ds^2 = H \Big( (dx^6)^2 + (dx^7)^2 \Big) 
+ H K^{-1} \Big( (dx^8)^2 + (dx^9)^2 \Big),
\notag \\
&
B = - \frac{A}{H^2 + A^2} dx^8 \wedge dx^9,
\quad
e^{2\phi} = H K^{-1},
\notag \\
&
K = H^2 + A^2.
\label{eq:522_solution}
\end{align}
It is obvious that the metric ceases to be a 
singlevalued 
function of
spacetime. Indeed, it is shown that the geometry has non-trivial
monodromy of the $O(D,D)$ transformation realized as the $\beta$-transformation.

The other important element of spacetime is the complex structure.
It is known that the type II NS5-brane geometry admits a bi-hypercomplex
structure.
In the current coordinate basis, each complex structure of the spacetime is
given by \cite{Papadopoulos:2000iv}
\begin{align}
&
J_{1,+} =
\left(
\begin{array}{cc}
0 & i \sigma_2 \\
i \sigma_2 & 0
\end{array}
\right),
&&
J_{1,-} =
\left(
\begin{array}{cc}
0 & - \sigma_1 \\
 \sigma_1 & 0
\end{array}
\right),
\notag \\
&
J_{2,+} = 
\left(
\begin{array}{cc}
0 & \mathbf{1}_2 \\
- \mathbf{1}_2 & 0
\end{array}
\right),
&&
J_{2,-} = 
\left(
\begin{array}{cc}
0 & \sigma_3 \\
- \sigma_3 & 0
\end{array}
\right),
\notag \\
&
J_{3,+} = 
\left(
\begin{array}{cc}
- i \sigma_2 & 0 \\
0 & i \sigma_2
\end{array}
\right),
&&
J_{3,-} = 
\left(
\begin{array}{cc}
- i \sigma_2 & 0 \\
0 & - i \sigma_2
\end{array}
\right).
\label{eq:bi-hypercomplex}
\end{align}
The corresponding fundamental two-forms are defined by $\omega_{a,\pm} = 
- g J_{a,\pm} = - H J_{a,\pm}$ ($a = 1,2,3$).
It is easy to show that $(J_{a,\pm}, \omega_{a,\pm})$ satisfies the
definition of the bi-hypercomplex structure.
Namely, $J_{a,+}$, $J_{a,-}$ commute with each other, satisfying the
SU(2) algebra independently and they are integrable.
The two-forms $\omega_{a,\pm}$ are covariantly constant.
In the NS5-brane frame, the $B$-field does not contribute to 
$J_{a,\pm}$ and $\omega_{a,\pm}$ 
and they do not pick up any monodromy.

However, the complex structures \eqref{eq:bi-hypercomplex} incorporate the components of the metric
and the $B$-field after T-duality transformations \cite{Kimura2022dma}.
For example, we find that the complex structure of the $5^2_2$-brane geometry
is given by
\begin{align}
&
J_{1,+} =
\left(
\begin{array}{cc}
0 & - HK^{-1} i \sigma_2 + A K^{-1} \mathbf{1}_2 \\
- H i \sigma_2 - A \mathbf{1}_2 & 0
\end{array}
\right),
\notag \\
&
J_{2,+} = 
\left(
\begin{array}{cc}
0 & - H K^{-1} \mathbf{1}_2 - A K^{-1} i \sigma_2 \\
H \mathbf{1}_2 - A i \sigma_2 & 0
\end{array}
\right),
\notag \\
&
J_{3,+} = 
\left(
\begin{array}{cc}
- i \sigma_2 & 0 \\
0 & i \sigma_2
\end{array}
\right),
\notag \\
&
J_{1,-} =
\left(
\begin{array}{cc}
0 & - H K^{-1} \sigma_1 - A K^{-1} \sigma_3 \\
H \sigma_1 + A \sigma_3 & 0
\end{array}
\right),
\notag \\
&
J_{2,-} = 
\left(
\begin{array}{cc}
0 & H K^{-1} \sigma_3 - A K^{-1} \sigma_1 \\
- H \sigma_3 + A \sigma_1 & 0
\end{array}
\right),
\notag \\
&
J_{3,-} = 
\left(
\begin{array}{cc}
- i \sigma_2 & 0 \\
0 & - i \sigma_2
\end{array}
\right).
\label{eq:522_bi-hypercomplex}
\end{align}
We also define the fundamental two-forms $\omega_{a,\pm} = - g
J_{a,\pm}$. 
Obviously
they lose their singlevaluedness.
Therefore there appears nontrivial monodromy in the bi-hypercomplex structure
$(J_{a,\pm}, \omega_{a,\pm})$ in the T-fold.
In the following section, we show that the monodromy becomes apparent in
the doubled geometry.

\section{Doubled formalism and complex geometry} \label{sec:doubled_formalism}

Double field theory (DFT) is a T-duality covariant gravity theory
defined on a $2D$-dimensional doubled space $\mathcal{M}$ \cite{Hull:2009mi}.
The physical degrees of freedom of the $D$-dimensional spacetime are
obtained by solving the strong constraint.
Generalized geometry \cite{Gualtieri2004}, which also makes manifest
T-duality, has a deep connection to doubled space.
For example, endomorphisms on $T\mathcal{M}$ and those
in generalized geometry are identified through the natural isomorphism
under the strong constraint \cite{Vaisman:2012ke, Freidel:2017yuv, Freidel:2018tkj}.
In the following, we assume that the strong constraint is always
imposed, which indicates that $T \mathcal{M}$ and the generalized tangent bundle are not distinguished.

We first provide the necessary tools for further discussion.
The generalized metric and the generalized dilaton are given by
\begin{align}
\mathcal{H}_{MN} = 
\left(
\begin{array}{cc}
g_{\mu \nu} - B_{\mu \rho} g^{\rho \sigma} B_{\sigma \nu} & B_{\mu \rho}
 g^{\rho \nu} \\
- g^{\mu \rho} B_{\rho \nu} & g^{\mu \nu}
\end{array}
\right),
\quad
e^{-2d} = \sqrt{-g} e^{-2\phi},
\end{align}
where $(g,B,\phi)$ are the $D$-dimensional spacetime metric, the $B$-field
and the dilaton.
The doubled indices $M, N, = 1, \ldots, 2D$ are raised and lowered by
the $O(D,D)$ invariant metric $\eta_{MN}$ and its inverse $\eta^{MN}$.
The factorized T-duality transformation $T_k$ along the isometry direction $x^k$
is given by 
\begin{align}
&
(h_k)^M {}_N 
=
\left(
\begin{array}{cc}
1 - t_k & t_k \\
t_k & 1 - t_k
\end{array}
\right)
\in O(D,D),
\notag \\ 
&
t_k = \mathrm{diag} (0, \ldots, 0, \overset{k}{\check{1}}, 0, \ldots, 0).
\end{align}
Together with the transformation of the doubled coordinate $x^M \to
(h_k)^M {}_N x^N$,
this $O(D,D)$ matrix acts on $\mathcal{H}_{MN}$ and 
an endomorphism 
$\mathcal{A}^M {}_N$
on $T \mathcal{M}$, which we call the doubled structure in the following,
as
\begin{align}
&
\mathcal{H}_{MN} \, \xrightarrow{T_k} \, \mathcal{H}'_{MN} = (h_k^t)_M {}^P
 \mathcal{H}_{PQ} (h_k)^Q {}_N,
\notag \\
&
\mathcal{A}^M {}_N \, \xrightarrow{T_k} \, \mathcal{A}^{\prime M} {}_N =
 (h_k^{-1})^M {}_P \mathcal{A}^P {}_Q (h_k)^Q {}_N.
\end{align}

Of particular importance is the following endomorphisms;
\begin{align}
&
(\Omega_{\Lambda})^M {}_N = 
\left(
\begin{array}{cc}
\Lambda^t & 0 \\
0 & \Lambda^{-1}
\end{array}
\right),
\quad
(e^B)^M {}_N = 
\left(
\begin{array}{cc}
1 & 0 \\
- B & 1
\end{array}
\right),
\notag \\
&
(e^{\beta})^M {}_N =
\left(
\begin{array}{cc}
1 & - \beta \\
0 & 1
\end{array}
\right),
\end{align}
where $\Lambda^{\mu} {}_{\nu} \in GL(D)$ is the 
$D$-dimensional general coordinate transformation, 
$B, \beta$ are $D$-dimensional anti-symmetric matrices which represent
the shifts of $B_{\mu \nu}, \beta^{\mu\nu}$ fields.

Notice that the generalized metric for $B \not= 0$ is obtained by 
$\mathcal{H}^0$ for $B=0$ by the $B$-transformation;
\begin{align}
\mathcal{H}^0 \, \xrightarrow{B} \,
\mathcal{H} =& \
(e^B)^t \mathcal{H}^0 (e^B)
\notag \\
=& \ 
\left(
\begin{array}{cc}
1 & B \\
0 & 1
\end{array}
\right)
\left(
\begin{array}{cc}
g & 0  \\
0 & g^{-1}
\end{array}
\right)
\left(
\begin{array}{cc}
1 & 0  \\
- B & 1
\end{array}
\right).
\end{align}
The same is true for any doubled structures
$ \mathcal{A}^0 \, \xrightarrow{B} \, \mathcal{A} = e^{-B} \mathcal{A}^0 e^B$.

Now we are ready to discuss branes of codimension two concretely.
In the following, we consider $D=4$.
The generalized metric for the NS5-brane solution
\eqref{eq:NS5_solution} is given by
\begin{align}
\mathcal{H}_{\mathrm{NS5}} (\theta) =& \
 (e^{B (\theta)})^t \mathcal{H}_{\mathrm{NS5}}^0 (e^{B (\theta)}),
\end{align}
where 
\begin{align}
\mathcal{H}_{\mathrm{NS5}}^0 = 
\left(
\begin{array}{cc}
H \mathbf{1}_4 & 0 \\
0 & H^{-1} \mathbf{1}_4
\end{array}
\right)
\end{align}
is the generalized metric for $B=0$ and 
\begin{align}
e^{B(\theta)} = 
\left(
\begin{array}{cc|cc}
\mathbf{1}_2 & 0 & 0 & 0 \\
0 & \mathbf{1}_2 & 0 & 0 \\
\hline
0 & 0 & \mathbf{1}_2 & 0 \\
0 & \sigma \theta \boldsymbol{\epsilon}_2 & 0 & \mathbf{1}_2
\end{array}
\right)
\label{eq:B-transf_mat}
\end{align}
is the $B$-field transformation.
From this expression, it is obvious that the NS5-brane of codimension two 
has a monodromy of the $B$-transformation;
\begin{align}
\mathcal{H}_{\mathrm{NS5}} (\theta = 2 \pi) = 
(e^{B (2\pi)})^t \mathcal{H}_{\mathrm{NS5}} (\theta = 0) (e^{B (2 \pi)}).
\end{align}

\subsection{T-duality transformations}

The T-duality transformation of the NS5-brane of codimension two yields the
KK5-brane of the same codimension.
This is obtained by the $O(D,D)$ transformation of the generalized metric;
\begin{align}
\mathcal{H}_{\mathrm{NS5}} \, \xrightarrow{T_9} \, 
\mathcal{H}_{\mathrm{KK5}}
 =& \ 
h_9^t (e^{B(\theta)})^t \mathcal{H}^0_{\mathrm{NS5}} (e^{B(\theta)}) h_9.
\end{align}
We note that this is rewritten as 
\begin{align}
\mathcal{H}_{\mathrm{KK5}} =& \  
h_9^t (e^{B(\theta)})^t (h_9^t)^{-1} 
h_9^t \mathcal{H}^0_{\mathrm{NS5}} h_9 \, 
h_9^{-1} (e^{B(\theta)}) h_9
\notag \\
=& \ 
\Big(
h_9^{-1} e^{B(\theta)} h_9
\Big)^t
\mathcal{H}_{\mathrm{KK5}}^0
\Big(
h_9^{-1} e^{B(\theta)} h_9
\Big).
\end{align}
Here we have defined the $\theta$-independent part of the 
generalized metric for the KK5-brane;
\begin{align}
\mathcal{H}_{\mathrm{KK5}}^0 =& \ 
h_9^t \mathcal{H}_{\mathrm{NS5}}^0 h_9
\notag \\
=& \ 
\left(
\begin{array}{cccc|cccc}
H & & & & & & & \\
 & H & & & & & & \\
 & & H & & & & & \\
 & & & H^{-1} & & & & \\
\hline
 & & & & H^{-1} & & & \\
 & & & & & H^{-1} & & \\
 & & & & & & H^{-1} & \\
 & & & & & & & H
\end{array}
\right).
\end{align}
On the other hand, the T-duality transformation of the
$B$-transformation is given by
\begin{align}
h_9^{-1} e^{B} h_9 =& \ 
\left(
\begin{array}{cccc|cccc}
1 & & & & & & & \\
 & 1 & & & & & & \\
 & & 1 & & & & & \\
 & & - \sigma \theta & 1 & & & & \\
\hline
 & & & & 1 & & & \\
 & & & & & 1 & & \\
 & & & & & & 1 & \sigma \theta \\
 & & & & & & & 1
\end{array}
\right)
\notag \\
=& \ 
\left(
\begin{array}{cc}
\Lambda^t & 0 \\
0 & \Lambda^{-1}
\end{array}
\right) = \Omega_{\Lambda (\theta)}.
\end{align}
Therefore we have
\begin{align}
\mathcal{H}_{\mathrm{KK5}} (2 \pi) = \Omega_{\Lambda (2\pi)}^t
\mathcal{H}_{\mathrm{KK5}} (\theta = 0) \Omega_{\Lambda (2\pi)}.
\end{align}
Then the monodromy of the KK5-brane is represented by the $O(D,D)$ matrix $\Omega_{\Lambda}$
that makes a $D$-dimensional general coordinate transformation.
This means that the monodromy of the KK5-brane is given by the general
coordinate transformation.

Applying the T-duality transformation $T_8$ to the KK5-brane yields the
$5^2_2$-brane.
The generalized metric is given by
\begin{align}
\mathcal{H}_{5^2_2} = h_8^t \mathcal{H}_{\mathrm{KK5}} h_8.
\end{align}
This is again rewritten as 
\begin{align}
\mathcal{H}_{5^2_2} = 
\Big\{
(
h_9 h_8
)^{-1} 
e^{B(\theta)} 
(
h_9 h_8
)
\Big\}^t
\mathcal{H}_{5^2_2}^0
\Big\{
(
h_9 h_8
)^{-1} 
e^{B(\theta)} 
(
h_9 h_8
)
\Big\}.
\end{align}
Here we have defined the $\theta$-independent part of the generalized
metric for the $5^2_2$-brane;
\begin{align}
\mathcal{H}^0_{5^2_2} 
=
\left(
\begin{array}{cccc|cccc}
H & & & & & & & \\
 & H & & & & & & \\
 & & H^{-1} & & & & & \\
 & & & H^{-1} & & & & \\
\hline
 & & & & H^{-1} & & & \\
 & & & & & H^{-1} & & \\
 & & & & & & H & \\
 & & & & & & & H
\end{array}
\right).
\end{align}
The $T_8 T_9$ transformation of the $B$-transformation is found to be
\begin{align}
(h_9 h_8)^{-1} e^B (h_9 h_8) 
=& \ 
\left(
\begin{array}{cccc|cccc}
1 & & & & & & & \\
 & 1 & & & & & & \\
 & & 1 & & & & & \sigma \theta \\
 & & & 1 & & & - \sigma \theta & \\
\hline
 & & & & 1 & & & \\
 & & & & & 1 & & \\
 & & & & & & 1 & \\
 & & & & & & & 1 
\end{array}
\right)
\notag \\
=& \ 
\left(
\begin{array}{cc}
1 & - \beta \\
0 & 1
\end{array}
\right)
=
e^{\beta (\theta)}.
\end{align}
Then we find that the monodromy of the $5^2_2$-brane is given by the
$\beta$-transformation,
\begin{align}
\mathcal{H}_{5^2_2} (2 \pi) = (e^{\beta (2\pi)})^t \mathcal{H}_{5^2_2}^0
 e^{\beta (2 \pi)}.
\end{align}

We note that these arguments can be generalized to heterotic theories.
Heterotic supergravities are formulated in the $O(D,D+n)$ gauged DFT
\cite{Hohm:2011ex}.
The generalized metric in the gauged DFT is given by
\begin{align}
\mathcal{H}_{MN} = 
\scalebox{.9}{$
\left(
\begin{array}{ccc}
g_{\mu \nu} + c_{\rho \mu} g^{\rho \sigma} c_{\sigma \nu}
+ A^2_{\mu \nu} & - g^{\nu \rho} c_{\rho \mu} & c_{\rho \mu} g^{\rho
\sigma} A_{\sigma \beta} \\
- g^{\mu \rho} c_{\rho \nu} & g^{\mu \nu} & - g^{\mu \rho} A_{\rho \beta}
\\
A_{\alpha \rho} g^{\rho \sigma} c_{\sigma \nu} + A_{\alpha \nu}
& - A_{\alpha \rho} g^{\nu \rho} & \kappa_{\alpha \beta} + A_{\alpha \rho} g^{\rho \sigma}
A_{\sigma \beta}
\end{array}
\right)
$},
\label{eq:gauged_generalized_metric}
\end{align}
where, 
after the gauging procedure, 
$A_{\mu \alpha}$, $\kappa_{\alpha \beta} \, (\alpha,\beta=1,\ldots,n)$ are
the gauge field and the Cartan-Killing metric for a gauge group $G \subset
O(D,D+n)$. 
Here we have defined
\begin{align}
c_{\mu \nu} = B_{\mu \nu} + \frac{1}{2} A^2_{\mu \nu}, 
\qquad
A^2_{\mu \nu} = \kappa_{\alpha \beta} A_{\mu \alpha} A_{\nu \beta}.
\end{align}
Analogous to the type II case, the generalized metric is decomposed as
\begin{align}
\mathcal{H} = (e^B)^t 
V^t
\mathcal{H}^0 
V (e^B),
\end{align}
where $\mathcal{H}^0$ is the generalized metric
\eqref{eq:gauged_generalized_metric} for $B_{\mu \nu} = A_{\mu \alpha} = 0$.
The matrix $e^B$ is an $O (D,D+n)$ generalization of the $B$-transformation
and $V$ may be called the $A$-transformation and is given by
\begin{align}
V^M {}_N = 
\left(
\begin{array}{ccc}
\delta^{\mu} {}_{\nu} & 0 & 0 \\
- \frac{1}{2} A^2_{\mu \nu} & \delta_{\mu} {}^{\nu} & - A_{\mu \beta} \\
A^{\alpha} {}_{\nu} & 0 & \delta^{\alpha} {}_{\beta}
\end{array}
\right).
\end{align}
For the NS5-brane of symmetric type in heterotic theories
\cite{Callan:1991at}, we have the codimension two solution given in
\eqref{eq:NS5_solution} together with the self-dual SU(2) gauge field in
the 't Hooft gauge;
\begin{align}
A_{\mu} = \bar{\sigma}_{\mu \nu} \del_{\nu} \log H (r).
\end{align}
Since the $V$ part does not introduce any product
$V e^{\text{const}.}$
in $\theta \to 2\pi$ for this solution, the monodromy of the symmetric solution in
heterotic theories is given by the $B$-transformation.
We can discuss the T-duality transformations of the five-branes of
codimension two in heterotic theories \cite{Sasaki:2016hpp,Sasaki:2017yrs}. 
This analysis implies that the symmetric
$5^2_2$-brane in heterotic theories has the monodromy of the
$\beta$-transformation, hence it is a T-fold.

We stress that the above discussion is based on the facts that 
(i) only the $B$-field is a linear function of $\theta$ in the NS5-brane frame,
(ii) the $B$-field is factored out as the $B$-transformation in the doubled formalism,
(iii) the T-duality transformations of the $B$-transformation result in
the $\Omega_{\Lambda}$ and the $\beta$ transformations.
These are 
properties
 peculiar to the codimension two solutions.

We have discussed only the metric and the $B$-field. 
In the following, we study the complex structures of the geometries.

\subsection{Generalized complex structures}

The above discussion on the monodromy property is inherited to the complex structures.

In order to discuss the T-duality transformation of complex structures, 
we introduce the following generalized complex structures
\cite{Gualtieri2004, Bredthauer:2006sz};
\begin{align}
\mathcal{J}_{J_{a,\pm}} = 
\left(
\begin{array}{cc}
J_{a,\pm} & 0 \\
0 & J^t_{a,\pm}
\end{array}
\right),
\qquad
\mathcal{J}_{\omega_{a,\pm}}
= 
\left(
\begin{array}{cc}
0 & - \omega^{-1}_{a,\pm} \\
\omega_{a,\pm} & 0
\end{array}
\right).
\end{align}
It is useful to consider the following combinations;
\begin{align}
\mathcal{J}_{a,\pm}^0
=& \ 
\frac{1}{2}
\Big[
(
\mathcal{J}_{J_{a,+}}
+
\mathcal{J}_{\omega_{a,+}}
)
\pm
(
\mathcal{J}_{J_{a,-}}
-
\mathcal{J}_{\omega_{a,-}}
)
\Big]
\notag \\
=& \ 
\frac{1}{2}
\left(
\begin{array}{cc}
J_{a,+} \pm J_{a,-}& - (\omega_{a,+}^{-1} \mp \omega_{a,-}^{-1}) \\
\omega_{a,+} \mp \omega_{a,-} & J^t_{a,+} \pm J^t_{a,-}
\end{array}
\right).
\end{align}
This parametrization remains the same after the T-duality transformations.
In addition, we stress that the $B$-field must be included in order to
perform the correct T-duality transformations of the complex structures \cite{Kimura2022dma}.
This is incorporated by the $B$-transformation of the generalized complex structures;
\begin{align}
\mathcal{J}_{a,\pm} = e^{-B} \mathcal{J}_{a,\pm}^0 e^B.
\label{eq:generalized_complex_B}
\end{align}
The T-duality transformations for the bi-hypercomplex structures of
the spacetime are performed by the $O(D,D)$ transformation $h_k$ on \eqref{eq:generalized_complex_B}.
As in the case of the generalized metric, in the KK5-brane frame we have
\begin{align}
\mathcal{J}'_{a,\pm}
= 
\Big( 
h_9^{-1} e^B h_9
\Big)^{-1}
\mathcal{J}_{a,\pm}^{\prime 0}
\Big( 
h_9^{-1} e^B h_9
\Big)
= \Omega_{\Lambda}^{-1} 
\mathcal{J}_{a,\pm}^{\prime 0}
\Omega_{\Lambda}.
\end{align}
Here we have again defined the $\theta$-independent part of the
generalized complex structure for the KK5-brane;
\begin{align}
\mathcal{J}_{a,\pm}^{\prime 0} = h_9^{-1} \mathcal{J}_{a,\pm}^{0} h_9.
\end{align}
Indeed, from this structure we can reconstruct the hyperK\"ahler structure
and the symplectic form of the Taub-NUT spacetime.
By this expression, it is obvious that the hyperK\"ahler structure of the
KK5-brane picks up the $\Omega_{\Lambda}$ monodromy. 

A further $T_8$ transformation yields the generalized complex structure
of the $5^2_2$-brane;
\begin{align}
\mathcal{J}''_{a,\pm}
=& \ 
\Big( 
(h_9 h_8)^{-1} e^{B (\theta)} (h_9 h_8)
\Big)^{-1}
\mathcal{J}_{a,\pm}^{\prime \prime 0}
\Big( 
(h_9 h_8)^{-1} e^{B (\theta)} (h_9 h_8)
\Big)
\notag \\
=& \ 
e^{-\beta (\theta)} 
\mathcal{J}_{a,\pm}^{\prime \prime 0} 
e^{\beta (\theta)}.
\label{eq:generalized_complex_522}
\end{align}
Here 
$\mathcal{J}^{\prime \prime 0}_{a,\pm}$ 
is the $\theta$-independent part of the 
generalized complex structure for the $5^2_2$-brane.
We can reconstruct the bi-hypercomplex structure of spacetime
\eqref{eq:522_bi-hypercomplex} from \eqref{eq:generalized_complex_522}. 
This is no longer singlevalued in
$D$-dimensions but this non-singlevaluedness is a reflection of the 
 monodromy of the $\beta$-transformation in the expression \eqref{eq:generalized_complex_522}.

\section{Monodromies of Connections and curvatures} \label{sec:codim_two_connection}
The monodromy nature appears also in other geometric quantities such as
 connections and curvatures.
The covariant derivative of the generalized general coordinate
transformation is defined as
\cite{,Jeon:2011cn, Hohm:2011si}
\begin{align}
\nabla_M V^N = \del_M V^N + \Gamma_{MP} {}^N V^P,
\end{align}
where $\Gamma_{MP} {}^N$ is an affine connection of $\mathcal{M}$.
The curvature $R_{MNK} {}^L$ and the torsion $T_{MN} {}^L$ are defined by the commutator of $\nabla_M$;
\begin{align}
[\nabla_M, \nabla_N] V_K = - R_{MNK} {}^L V_L - T_{MN} {}^L \nabla_L
 V_K,
\label{eq:curvature_def}
\end{align}
where $T_{MN} {}^L = 2 \Gamma_{[MN]} {}^L$.
Since this is {\it not} covariant with respect to the {\it generalized} general
coordinate transformation, we introduce a covariant torsion given by
\cite{,Jeon:2011cn, Hohm:2011si}
\begin{align}
\mathcal{T}_{MN} {}^L = T_{MN} {}^L + \eta^{LP} \eta_{NQ} \Gamma_{PM} {}^Q.
\end{align}
This is called the generalized torsion tensor.
The doubled space is assumed to satisfy $\mathcal{T}_{MN} {}^L = 0$.
Requiring the consistency with $\mathcal{H}_{MN}$, $\eta_{MN}$ and $d$, 
the connection is given as
\begin{align}
\Gamma_{MNK} = \hat{\Gamma}_{MNK} + \Sigma_{MNK},
\end{align}
where $\Sigma_{MNK}$ is the undetermined part satisfying the condition $\eta^{MK} \Sigma_{MNK} = 0$.
The determined part is given by 
\begin{align}
\hat{\Gamma}_{MNK} =& \ 
- 2 (P \del_M P)_{[NK]}
- 2 
\Big\{
\bar{P}_{[N} {}^P \bar{P}_{K]} {}^Q
-
P_{[N} {}^P P_{K]} {}^Q
\Big\} \del_P P_{QM}
\notag \\
& \ 
+ \frac{4}{D-1}
\Big\{
P_{M[N} P_{K]} {}^Q
+
\bar{P}_{M[N} \bar{P}_{K]} {}^Q
\Big\} 
\Big\{
\del_Q d
+
(P \del^P P)_{[PQ]}
\Big\},
\label{eq:conn_determined_part}
\end{align}
where the projection operators are defined as 
\begin{align}
P_{MN} = \frac{1}{2} (\eta_{MN} - \mathcal{H}_{MN}),
\qquad
\bar{P}_{MN} = \frac{1}{2} (\eta_{MN} + \mathcal{H}_{MN}).
\label{eq:projection_operators}
\end{align}

From the definition \eqref{eq:curvature_def}, the curvature is defined as 
\begin{align}
R_{MNK} {}^L = \del_M \Gamma_{NK} {}^L - \del_N \Gamma_{MK} {}^L +
 \Gamma_{MQ} {}^L \Gamma_{NK} {}^Q - \Gamma_{NQ} {}^L \Gamma_{MK} {}^Q.
\end{align}
Since this is neither covariant with respect to the generalized general
coordinate transformation, we introduce the covariantized curvature
$\mathcal{R}_{MNKL}$ defined by 
\begin{align}
\mathcal{R}_{MNKL} = R_{MNKL} + R_{KLMN} + \Gamma_{QMN} \Gamma^Q {}_{KL}.
\end{align}
This is called the generalized Riemann curvature tensor.
Then the generalized Ricci tensor is defined by
\begin{align}
\mathcal{R}_{MN} = \eta^{KL}
\Big(
P_K {}^P P_M {}^Q \bar{P}_N {}^R P_L {}^S
+
P_K {}^P \bar{P}_N {}^Q P_M {}^R P_L {}^S
\Big)
 \mathcal{R}_{PQRS}.
\end{align}
Similarly, the generalized Ricci scalar is defined by 
\begin{align}
\mathcal{R} = P^{MN} \mathcal{R}_{MN}.
\end{align}

Although the undetermined part $\Sigma_{MNK}$ 
cannot be precluded only by the covariance, 
the $\hat{\Gamma}_{MNP}$ part is uniquely determined \cite{Jeon:2011cn}.
In addition, since the undetermined part in the connection is projected out in the
Ricci curvature \cite{Hohm:2011si}, we focus only on the determined part
$\hat{\Gamma}_{MNP}$ in the following.

We now examine the relation between the $D$-dimensional geometric
quantities associated with the curvatures.
The generalized Riemann curvature tensor $\mathcal{R}_{MNKL}$ does
not contain the $D$-dimensional Riemann curvature tensor $R^{\mu}
{}_{\nu\rho\sigma}$ \cite{,Jeon:2011cn, Hohm:2011si}.
However, we find that 
the $D$-dimensional Ricci tensor $R_{\mu\nu}$ is embedded in $\mathcal{R}_{MN}$.
When we set $B= \phi = 0$ and substitute the generalized metric into
$\mathcal{R}_{MN}$, we have
\begin{align}
\mathcal{R}_{\mu\nu} &= \frac{1}{2} R_{\mu\nu}, \qquad
\mathcal{R}_\mu{}^\nu = 0, \qquad 
\mathcal{R}^{\mu\nu} = - \frac{1}{2} g^{\mu\alpha} g^{\nu\beta} R_{\alpha\beta}.
\end{align}
Therefore $\mathcal{R}_{MN}$ is regarded as a T-duality covariant
generalization of the $D$-dimensional curvature.

Although we have $\mathcal{R} = \mathcal{R}_{MN} = 0$ due to
the equation of motion of DFT, it is significant to study the
monodromy nature of the connection and the curvature. 
To this end, we first study the
monodromy of the doubled derivative of the generalized metric
$\partial_M \mathcal{H}_{NK}$.
We focus on the type II NS5-brane solution.
Since the generalized metric for the NS5-brane depends only on
$r$ and $\theta$, $\mathcal{H}_{\text{NS5}} = \mathcal{H}_{\text{NS5}}
(r, \theta)$, everything except the $r$- and the $\theta$-derivatives vanishes trivially.
Therefore, we only need to consider $\partial_r \mathcal{H}_{\text{NS5}}$ and $\partial_\theta \mathcal{H}_{\text{NS5}}$.
The $r$-derivative of the NS5 generalized metric is
\begin{align}
\partial_r \mathcal{H}_{\text{NS5}} 
&= (e^{B(\theta)})^t (\partial_r \mathcal{H}_{\text{NS5}}^0) (e^{B(\theta)}).
\end{align}
On the other hand, the $\theta$-derivative of $\mathcal{H}_{\text{NS5}}$ is given by
\begin{align}
\partial_\theta \mathcal{H}_{\text{NS5}} 
&= (\partial_\theta (e^{B(\theta)})^t) \mathcal{H}_{\text{NS5}}^0 e^{B(\theta)}
+ (e^{B(\theta)})^t \mathcal{H}_{\text{NS5}}^0 (\partial_\theta e^{B(\theta)}).
\end{align}
Now the derivative of $e^{B(\theta)}$ is calculated as 
\begin{align}
\partial_\theta e^{B(\theta)} =& \ 
\left(
\begin{array}{cc|cc}
0 & 0 & 0 & 0 \\
0 & 0 & 0 & 0 \\
\hline
0 & 0 & 0 & 0 \\
0 & \sigma \boldsymbol{\epsilon}_2 & 0 & 0
\end{array}
\right), 
\notag \\
\partial_\theta (e^{B(\theta)})^t =& \ 
\left(
\begin{array}{cc|cc}
0 & 0 & 0 & 0 \\
0 & 0 & 0 & - \sigma \boldsymbol{\epsilon}_2 \\
\hline
0 & 0 & 0 & 0 \\
0 & 0 & 0 & 0
\end{array}
\right).
\label{eq:deriv_B-transf_mat}
\end{align}
Using this and the $B$-transformation matrix \eqref{eq:B-transf_mat}, we find
\begin{align}
&
(\partial_\theta e^{B(\theta)}) e^{-B(\theta)} 
\notag \\
&= 
\left(
\begin{array}{cc|cc}
0 & 0 & 0 & 0 \\
0 & 0 & 0 & 0 \\
\hline
0 & 0 & 0 & 0 \\
0 & \sigma \boldsymbol{\epsilon}_2 & 0 & 0
\end{array}
\right)
\left(
\begin{array}{cc|cc}
\mathbf{1}_2 & 0 & 0 & 0 \\
0 & \mathbf{1}_2 & 0 & 0 \\
\hline
0 & 0 & \mathbf{1}_2 & 0 \\
0 & - \sigma \theta \boldsymbol{\epsilon}_2 & 0 & \mathbf{1}_2
\end{array}
\right)
\notag \\
&= 
\left(
\begin{array}{cc|cc}
0 & 0 & 0 & 0 \\
0 & 0 & 0 & 0 \\
\hline
0 & 0 & 0 & 0 \\
0 & \sigma \boldsymbol{\epsilon}_2 & 0 & 0
\end{array}
\right)
= \partial_\theta e^{B(\theta)}.
\end{align}
Similarly we have
$(e^{-B(\theta)})^t \partial_\theta (e^{B(\theta)})^t = \partial_\theta
(e^{B(\theta)})^t$.
With these results at hand and the fact 
$e^{-B(\theta)} e^{B(\theta)} = 1 = (e^{B(\theta)})^t (e^{-B(\theta)})^t$,
the $\theta$-derivative of the generalized metric for the NS5-brane is 
\begin{align}
\partial_\theta \mathcal{H}_{\text{NS5}} 
&= (e^{B(\theta)})^t \big\{ 
	(\partial_\theta (e^{B(\theta)})^t) \mathcal{H}_{\text{NS5}}^0 
	+ \mathcal{H}_{\text{NS5}}^0 (\partial_\theta e^{B(\theta)}) 
	\big\} e^{B(\theta)}
\notag \\
&= (e^{B(\theta)})^t 
\left(
\begin{array}{cc|cc}
0 & 0 & 0 & 0 \\
0 & 0 & 0 & - H^{-1} \sigma \boldsymbol{\epsilon}_2 \\
\hline
0 & 0 & 0 & 0 \\
0 & H^{-1} \sigma \boldsymbol{\epsilon}_2 & 0 & 0
\end{array}
\right) e^{B(\theta)}
\notag \\
&= (e^{B(\theta)})^t 
(\partial_\theta \mathcal{H}_{\text{NS5}}^0)
e^{B(\theta)}.
\end{align}

Notice that $\partial_\theta \mathcal{H}_{\text{NS5}}^0$ is not the
$\theta$-derivative of $\mathcal{H}_{\text{NS5}}^0$ but $\partial_\theta
\mathcal{H}_{\text{NS5 }} \big|_{\theta = 0}$.
This $\partial_\theta \mathcal{H}_{\text{NS5 }}^0$ is a matrix that depends only on $r$.
Also, the doubled derivative for a function $f = f(r,\theta)$, which
in general depends only on $r$ and $\theta$, satisfies the following relation;
\begin{align}
&
(e^{B(\theta)})^N{}_M \partial_N f
\notag \\
&= \scalebox{.8}{$
\left(
\begin{array}{cccc|cccc}
1 & 0 & 0 & 0 & 0 & 0 & 0 & 0 \\
0 & 1 & 0 & 0 & 0 & 0 & 0 & 0 \\
0 & 0 & 1 & 0 & 0 & 0 & 0 & -\sigma \theta \\
0 & 0 & 0 & 1 & 0 & 0 & \sigma \theta & 0 \\ \hline
0 & 0 & 0 & 0 & 1 & 0 & 0 & 0 \\
0 & 0 & 0 & 0 & 0 & 1 & 0 & 0 \\
0 & 0 & 0 & 0 & 0 & 0 & 1 & 0 \\
0 & 0 & 0 & 0 & 0 & 0 & 0 & 1 
\end{array}
\right) 
\begin{pmatrix}
\partial_r f \\
\partial_\theta f \\
0 \\
0 \\
0 \\
0 \\
0 \\
0
\end{pmatrix}$}
= \scalebox{.8}{$
\begin{pmatrix}
\partial_r f \\
\partial_\theta f \\
0 \\
0 \\
0 \\
0 \\
0 \\
0
\end{pmatrix}$}
= \partial_M f.
\label{eq:monodromy_doubled_deriv}
\end{align}
Thus, the derivative of the generalized metric for the NS5-brane has the following structure;
\begin{align}
\partial_M (\mathcal{H}_{\text{NS5}})_{NK} 
&= (e^{B(\theta)})^P{}_M (e^{B(\theta)})^Q{}_N (e^{B(\theta)})^R{}_K \partial_P (\mathcal{H}_{\text{NS5}}^0)_{QR}.
\end{align}
Let us also examine the monodromy of the second-order derivative $\partial_M \partial_N \mathcal{H}_{KL}$.
The second-order $r$-derivative of the NS5 generalized metric is given by
\begin{align}
\partial_r^2 \mathcal{H}_{\text{NS5}} 
&= (e^{B(\theta)})^t (\partial_r^{2} \mathcal{H}_{\text{NS5}}^0) e^{B(\theta)}.
\end{align}
Similarly, the second-order $\theta$-derivative is calculated to be
\begin{align}
\partial_\theta^2 \mathcal{H}_{\text{NS5}} 
&= 2 (\partial_\theta (e^{B(\theta)})^t) \mathcal{H}_{\text{NS5}}^0 (\partial_\theta e^{B(\theta)})
\notag \\
&= (e^{-B(\theta)})^t \big\{ 
	2 (\partial_\theta (e^{B(\theta)})^t) \mathcal{H}_{\text{NS5}}^0 (\partial_\theta e^{B(\theta)}) 
	\big\} e^{B(\theta)}.
\end{align}
Since the $\theta$-derivative of the $B$-transformation is a constant matrix,
$(\partial_\theta (e^{B(\theta)})^t) \mathcal{H}_{\text{NS5}}^0 (\partial_\theta e^{B(\theta)})$ is independent of $\theta$.
This matrix is given by calculating $\partial_\theta^2
\mathcal{H}_{\text{NS5}} (r,\theta)$ and then take $\theta = 0$.
Furthermore, $\partial_r \partial_\theta \mathcal{H}_{\text{NS5}}$ is
given by
\begin{align}
&
\partial_r \partial_\theta \mathcal{H}_{\text{NS5}} 
\notag \\
&= (\partial_\theta (e^{B(\theta)})^t) (\partial_r \mathcal{H}_{\text{NS5}}^0) e^{B(\theta)} 
+ (e^{B(\theta)})^t (\partial_r \mathcal{H}_{\text{NS5}}^0) (\partial_\theta e^{B(\theta)})
\notag \\
&= (e^{B(\theta)})^t \big\{ 
	(\partial_\theta (e^{B(\theta)})^t) (\partial_r \mathcal{H}_{\text{NS5}}^0)
	+ (\partial_r \mathcal{H}_{\text{NS5}}^0) (\partial_\theta
 e^{B(\theta)}) \big\} e^{B(\theta)}, 
\end{align}
where 
$(\partial_\theta (e^{B(\theta)})^t) (\partial_r \mathcal{H}_{\text{NS5}}^0)
+ (\partial_r \mathcal{H}_{\text{NS5}}^0) (\partial_\theta e^{B(\theta)})$
is a matrix independent of $\theta$ by the same reason as above.
This matrix is also equal to the matrix $\partial_r \partial_\theta
\mathcal{H}_{\text{NS5}} (r,\theta)$ calculated first and then take $\theta = 0$.
By the same argument as for the first-order derivative of the
generalized metric, the monodromy of the second-order derivative is given by
\begin{align}
&
\partial_M \partial_N (\mathcal{H}_{\text{NS5}})_{KL} 
\notag \\
&= (e^{B(\theta)})^P{}_M (e^{B(\theta)})^Q{}_N (e^{B(\theta)})^R{}_K
 (e^{B(\theta)})^S{}_L \partial_P \partial_Q
 (\mathcal{H}_{\text{NS5}}^0)_{RS}.
\end{align}
We next examine the monodromy of the determined part of the connection.
The determined part of the connection contains the derivatives of the projection operators.
The monodromy nature of this is calculated 
by using the derivative of the generalized metric;
\begin{align}
\partial_M P_{NK} 
&= - \frac{1}{2} \partial_M \mathcal{H}_{NK} 
\notag \\
&= - \frac{1}{2} (e^{B(\theta)})^P{}_M (e^{B(\theta)})^Q{}_N (e^{B(\theta)})^R{}_K \partial_P \mathcal{H}_{QR}^0 
\notag \\
&= (e^{B(\theta)})^P{}_M (e^{B(\theta)})^Q{}_N (e^{B(\theta)})^R{}_K \partial_P P_{QR}^0,
\end{align}
where $P^0_{QR}$ is the $\theta$-independent part of the projection
operator \eqref{eq:projection_operators}.
Since the generalized dilaton $d$ is a function of $r$ only, the result
\eqref{eq:monodromy_doubled_deriv} is also applicable.
Thus, the monodromy of the determined part of the connection is given by
\begin{align}
\hat{\Gamma}_{MNK} 
&= (e^{B(\theta)})^P{}_M (e^{B(\theta)})^Q{}_N (e^{B(\theta)})^R{}_K \hat{\Gamma}_{PQR}^0,
\label{eq:monodromy_conn_detd_part}
\end{align}
where again $\hat{\Gamma}^0_{PQR}$ is the $\theta$-independent part of
the connection.
We consider the monodromy of the determined part of the curvature.
The derivative of the determined part of the connection $\partial_M
\hat{\Gamma}_{NKL}$ includes the second-order derivative of the projection operator from \eqref{eq:conn_determined_part}.
Therefore, the monodromy of the derivative of the determined part of the
connection is 
\begin{align}
\partial_M \hat{\Gamma}_{NKL} 
&= (e^{B(\theta)})^P{}_M (e^{B(\theta)})^Q{}_N (e^{B(\theta)})^R{}_K (e^{B(\theta)})^S{}_L \partial_P \hat{\Gamma}_{QRS}^0.
\end{align}
Using this result, we find the monodromy of the determined part of the curvature is
\begin{align}
\hat{R}_{MNKL} 
&= (e^{B(\theta)})^P{}_M (e^{B(\theta)})^Q{}_N (e^{B(\theta)})^R{}_K (e^{B(\theta)})^S{}_L \hat{R}_{PQRS}^0.
\label{eq:monodromy_curv_detd_part}
\end{align}
Again, the superscript ``0'' means $\theta$-independent.
Similarly, we have 
\begin{align}
\hat{\mathcal{R}}_{MN} 
&= (e^{B(\theta)})^P{}_M (e^{B(\theta)})^Q{}_N \hat{\mathcal{R}}_{PQ}^0.
\end{align}

As with the other geometric structures, we can consider the monodromy of
the determined parts of connection and the curvatures in the KK5- and the $5^2_2$-frames.
We denote the determined part of the connection in the KK5-frame by
$\hat{\Gamma}_{MNK}'$ and the determined part of the curvature by $\hat{\mathcal{R}}_{MNKL}'$.
By applying the $T_9$-transformation to the connection and the
determined part of curvature in the NS5-frame, 
we can show that each monodromy in the KK5-frame is given by
\begin{align}
\hat{\Gamma}_{MNK}'
&= (\Omega_\Lambda)^P{}_M (\Omega_\Lambda)^Q{}_N (\Omega_\Lambda)^R{}_K
\hat{\Gamma}_{PQR}^{\prime 0},
\notag \\
\hat{\mathcal{R}}_{MNKL}'
&= (\Omega_\Lambda)^P{}_M (\Omega_\Lambda)^Q{}_N (\Omega_\Lambda)^R{}_K (\Omega_\Lambda)^S{}_L 
\hat{\mathcal{R}}_{PQRS}^{\prime 0}.
\end{align}
Furthermore, after the $T_8$-transformation, each monodromy in the
$5^2_2$-frame is given by
\begin{align}
\hat{\Gamma}_{MNK}''
&= (e^\beta)^P{}_M (e^\beta)^Q{}_N (e^\beta)^R{}_K 
\hat{\Gamma}_{PQR}^{\prime\prime 0},
\notag \\
\hat{\mathcal{R}}_{MNKL}''
&= (e^\beta)^P{}_M (e^\beta)^Q{}_N (e^\beta)^R{}_K (e^\beta)^S{}_L 
\hat{\mathcal{R}}_{PQRS}^{\prime\prime 0}.
\end{align}
Note that the transformation of the doubled coordinate $x^M \to (h_k)^M
{}_N x^N$ in the derivatives becomes trivial since we have $(h^{-1}_k)^N
{}_M \del_N f = \del_M f$ similar to the relation \eqref{eq:monodromy_doubled_deriv}.
This shows that the geometry of the T-fold is completely characterized by the
monodromy of the $\beta$-transformation.

\section{Conclusion} \label{sec:conclusion}

In this letter, we studied the spacetime geometries for branes of codimension two.
In particular, we focused on the NS5-, the KK5- and the $5^2_2$-branes in
type II and heterotic theories as prototypical examples.
These geometries are characterized by the monodromy associated with the
$B$-transformation, the general coordinate transformation, and the $\beta$-transformation. 
The geometries are best described in the $2D$-dimensional doubled space $\mathcal{M}$.
The various geometric quantities necessary for the description of
spacetime, i.e., the metric, the connection, the curvature, the complex
structures and the fundamental forms are all embedded into the doubled
(generalized) structures on the doubled space.
They are the generalized metric, the generalized connection, the
generalized curvature, and the generalized complex structure.
We showed that the monodromy nature of the geometries are
linearly represented by these generalized structures.
In particular, the $B$-, the general coordinate, and the
$\beta$-transformations are related by T-duality transformations. 
The manifestation of the $\beta$ monodromy in the T-fold becomes obvious
in doubled space.

In particular, in T-fold spacetimes, the curvature of spacetime becomes
essentially a multivalued function and loses its geometric meaning in
$D$ dimensions in the conventional manner.
However, this feature is revealed as the monodromy of the
$\beta$-transformation which is an automorphism in the $2D$-dimensional
doubled space.
The same is true even for the complex structures.
The geometry of the T-fold is clearly well-defined in the doubled space.
It would also be interesting to study the monodromy of the curvature and
the complex structures in the exceptional geometry
\cite{Aldazabal:2013mya, Cederwall:2013naa, Hassler:2023axp}.
We will come back to these issues in future studies.

\section*{Acknowledgements}
The authors would like to thank Yuho Sakatani for useful comments and information on
references.
The work is supported in part by Grant-in-Aid for Scientific Research
(C), JSPS KAKENHI Grant Numbers JP20K03952 (S.S. and T.K.) and JP23K03398 (T.K.).

\end{document}